\documentclass{llncs}
\usepackage[english]{babel}
\usepackage[latin1]{inputenc}
\usepackage{graphicx}
\usepackage{listings}
\usepackage{amsmath}
\usepackage{algorithm}
\usepackage{algorithmic}
\usepackage{amsfonts}
\usepackage{amssymb}

\title{Cardinality heterogeneities in Web~service composition: Issues and solutions}

\author{
M. Mrissa$^{1}$, Ph. Thiran$^{1}$, J-M. Jacquet$^{1}$,\\
D. Benslimane$^{2}$ and Z. Maamar$^{3}$}

\institute{
{\small $^{1}$University of Namur, Namur, Belgium}\\{\small \emph{\{michael.mrissa, philippe.thiran, jean-marie.jacquet\}@fundp.ac.be}}\\
{\small $^{2}$Claude Bernard Lyon 1 University, Lyon, France}\\{\small \emph{djamal.benslimane@liris.cnrs.fr}}\\
{\small $^{3}$Zayed University, Dubai, United Arab
Emirates}\\{\small \emph{zakaria.maamar@zu.ac.ae}}}

\begin{document}

\maketitle{}

\begin{abstract}
Data exchanges between Web services engaged in a composition raise several heterogeneities. In this paper, we address the problem of data cardinality heterogeneity in a composition. Firstly, we build a theoretical framework to describe different aspects of Web services that relate to data cardinality, and secondly, we solve this problem by developing a solution for cardinality mediation based on constraint logic programming.

\end{abstract}

\section{Introduction}\label{sec:introduction}
Web services are independent software components that users or other
peers can invoke in order to utilize their functionalities, like
\texttt{WeatherForecast} and \texttt{RoomBooking}. Web services
combine the benefits of service-oriented computing  paradigm and
platform-independent protocols~(HTTP~\cite{http}) to enable and
sustain business-to-business collaborations. To make these
collaborations happen and last for long periods of time, Web
services rely on a set of XML-based protocols and languages that
support their discovery (UDDI~\cite{uddi}), description
(WSDL~\cite{wsdl}) and invocation (SOAP~\cite{soap}).

Composition orchestrates the functionalities of several Web services
into the same loosely-coupled business processes to answer complex
users' needs. Different languages exist to specify composition scenarios
in terms of Web services to include, interactions to allow,
exceptions to handle, just to cite a few. \texttt{WS-BPEL} is
nowadays the \textit{defacto} composition standard~\cite{bpel}.
However despite this ``battery'' of protocols and languages,
composition remains a tedious task. Web services continue to be
designed in isolation from each other, which increases the levels of
heterogeneities between them. In today's economy, it is unlikely
that suppliers will develop the same types of Web services and
comply with the same design options.

In this paper, we look into these heterogeneities from a
data-cardinality perspective. Cardinality typically refers to the number of elements in a
set or group, and is considered as a property of that
grouping~(Wordnet~\cite{cardinality}). In the context of Web services composition,
we refer to cardinality as the number of data instances contained in
the messages that Web services engaged in composition exchange. Web
services have different \textit{limitations} in terms of minimum and
maximum data cardinalities, and this for several reasons such as technical limitations, search for interoperability with specific partners, quality of service depending on the number of results provided, etc. We refer to these limitations as \textit{cardinality constraints}.
Mismatching cardinality constraints will for sure hamper the smooth
progress of a composition by making Web services, for example,
indefinitely wait for the right number of elements or return invalid
responses because of the lack of appropriate elements. Additional
illustrative examples are provided throughout this paper.

While cardinality heterogeneities have been tackled in the field of
schema matching~\cite{avigdor05}, and despite the large amount of
work on Web services mediation, the resolution of cardinality
heterogeneities between Web services remains somehow marginalized.
In~\cite{DBLP:conf/icws/NagarajanVSML06}, Nagarajan~et~al. present a
classification of Web services heterogeneities. In spite of
an exhaustive classification, the authors do not explicitly mention
cardinality heterogeneities. Instead, they mention an ``entity
level'' category of data incompatibilities, to which cardinality
heterogeneities belong. In the following, we specifically focus on
cardinality heterogeneities, and assume that semantic and
structural data heterogeneities are already fixed.

The rest of this paper is structured as follows.
Section~\ref{sec:presentation} presents the vocabulary and
theoretical background that was developed to tackle the cardinality
concern. Section~\ref{sec:problem} lists the different cases of
cardinality heterogeneities that arise between Web services. Section~\ref{sec:card_mediation} describes
the proposed solution along with its theoretical and implementation framework, prior
to concluding in Section~\ref{sec:conclusion}.

\section{Theoretical framework}
\label{sec:presentation} Our work starts by defining various
concepts such as data flow, Web service, and composition. The
purpose of these definitions is to formalize the cardinality issue,
and provide a solid background for the proposed solution.

\subsection{Data flow representation}
A composition orchestrates several Web services into a business
process. In this process, Web services typically manage data and
exchange them with peers in compliance with some predefined flows.

\subsubsection{Characteristics of data flow:}
\begin{figure}[ht]
 \centering
 \includegraphics[width=.9\textwidth]{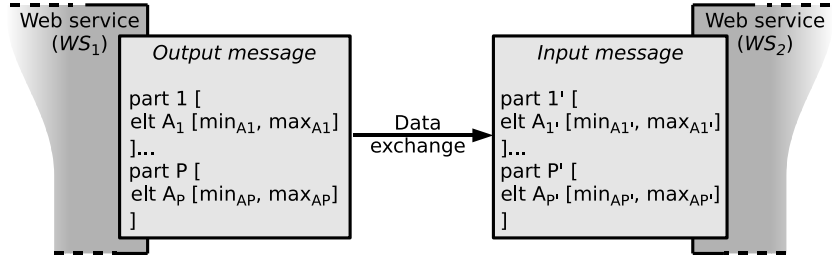}
 \caption{Simple data flow in a composition}
 \label{fig:data_flow}
\end{figure}

Fig.~\ref{fig:data_flow} illustrates a simple data flow between two
Web services: data are passed on from a source Web service~$WS_1$
(\textit{sender}) to a target Web service $WS_2$
(\textit{receiver}). These data are organized in terms of input and
output messages that are structured using several parts. Each
message part has a \textit{type} described with an~XML
Schema~\cite{xmlschema}. This \textit{type} may contain additional
elements integrated into a complex structure. Cardinality
constraints on data are expressed at the type level using
$minOccurs$ and $maxOccurs$ XML Schema attributes. Examples of such
constraints in Fig.~\ref{fig:data_flow} are [$min_{A1}$,$max_{A1}$],
[$min_{AP}$,$max_{AP}$], [$min_{A1'}$,$max_{A1'}$], and
[$min_{AP'}$,$max_{AP'}$].

For simplification purposes, this paper deals with messages that contain
one part with a data type that contains one couple of $minOccurs$ and
$maxOccurs$ cardinality constraints. However, this does not limit the
applicability of our solution to complex data structures and
multi-part messages.

\subsubsection{Constraints on data flows:}
\label{sec:users_req} In a composition, different constraints that relate to the cardinality concern apply to data flows. We illustrate these constraints
with the following examples:
\begin{itemize}
\item \textbf{Example $1$:} Editor $WS_1$ sends data to printing~$WS_2$.

\item \textbf{Example $2$:} Google-like~$WS_1$ sends data to
mashup~$WS_2$.

\item \textbf{Example $3$:} Shopping~$WS_1$ sends data for payment
to banking~$WS_2$.
\end{itemize}

\paragraph{Data selection constraint.} Denotes the
possibility of selecting some elements out of the data flow. In
Example~$1$, data selection is not authorized, i.e.,~all data
from~$WS_1$ must be printed. Similarly, all shopping items must be
processed by the banking Web service in Example~$3$. Data flows in
both examples are not data-selection tolerant. In Example~$2$, $WS_2$~only selects the first elements in the list, because search results are classified according to their relevance, and as a consequence the first results are the most relevant. If the mashup~$WS_2$ offers three entries for the search answers, only the first three results are selected. In this example, the data flow is data-selection tolerant.

\paragraph{Duplicate constraint.} Denotes whether
receiver Web services tolerate incoming data with duplicate
elements. In Examples~$1$ and~$3$, data flows are duplicate
tolerant. Indeed, a same document can be printed several times, and
a shopping item bought several times needs to be paid several times.
In Example~$2$, duplicates are not tolerated and should be merged
into a unique element, as users are not interested in duplicate
search results, so the data flow is not duplicate
tolerant.

\paragraph{Ordering constraint.} Relates to how much changes in
the order of elements in the data flow are accepted. Both
Data flows of examples~$1$ and~$2$ are not order-change tolerant, i.e.,~the order
of transmitted elements needs to be maintained. Indeed, the order of
search results is important to the user, and the order of printed
documents in also relevant to the printing Web service. In
Example~$3$, data flow is order-change tolerant, all the bought items
have the same priority and the payment order does not affect the
banking Web service.

The aforementioned three types of constraints permit describing data
flows along the following aspects: (i)~\textit{data selection}
attribute (boolean) denotes whether specific parts of data can be
selected, (ii)~\textit{duplicate} attribute (boolean) denotes
whether duplicate elements are tolerated in the data flow, and
(iii)~\textit{ordering} attribute (boolean) denotes whether the
order of the elements must be conserved. This classification of
constraints is particularly relevant during the
cardinality-mediation exercise (Section~\ref{sec:card_mediation}). Each attribute impacts the number and organization of elements that Web services exchange and the mediation to adopt
per type of constraint.

\subsection{Data schema and constraint representation}
To highlight the cardinality issue in the definition of data schema,
we follow the definition of a~\emph{schema graph} given in~\cite{Smiljanic04}. A schema graph is a labeled directed graph with property sets. In this graph, nodes represent element types, edges represent relationships, and property sets on nodes or edges describe specific XML~features. In this paper, we define a~\textit{constrained schema} as a schema graph with its cardinality constraints described via property sets, but we remind that property sets also describe other XML~features in the original work~\cite{Smiljanic04}.

\begin{definition}[Constrained Schema]
A \emph{constrained schema} is a tuple $CS = \langle ET, R, s, t,
PS\rangle$ where:
\begin{itemize}
\item[-] $ET$ is a nonempty finite set of element types;

\item[-] $R \subseteq (ET \times ET)$ is a finite set of relationships;

\item[-] $s: R \rightarrow ET$ is a function that indicates the source of a relationship;

\item[-] $t: R \rightarrow ET$ is a function that indicates the target of a relationship;

\item[-] $PS:\lbrace ET \cup R \rbrace \times P \rightarrow V$ is a
function that represents property sets, where $P$ is a set of
properties and $V$ is a set of values including the $null$ value.
\end{itemize}

In the context of XML Schema, $V = \mathbb{R} \cup \mathbb{S} \cup
\mathbb{U} \cup \lbrace \varnothing \rbrace$ where $\mathbb{R}$,
$\mathbb{S}$, $\mathbb{U}$ are sets of real numbers, strings, and
user-defined labels, respectively.
\end{definition}

\begin{definition}[Cardinality Constraint]
A {\em cardinality constraint} $k$ is a property of a constrained schema $CS = \langle ET, R, s, t, PS\rangle$, as aforementioned. This property is associated with a relationship $r$ of $CS$ and is
defined as follows:
\begin{description}
    \item[-] \textbf{minCard:} $minCard(r) = i$ where $i \in \aleph^+$;
    \item[-] \textbf{maxCard:} $maxCard(r) = j$ where $j \in \aleph^+$ and $i \leq j$.
\end{description}
\end{definition}
In the rest of this paper, a cardinality constraint on a
relation $r\in R$ is denoted as $k_r$ = [$i$, $j$].

\subsection{Web service representation}
We simply represent Web services as black boxes accepting inputs and
returning outputs. Additionally, we consider the possibility for a Web service to be invoked several times and to return additional results. Such invocation possibility can be exploited for the purpose of cardinality mediation.

\subsubsection{Maximum number of invocations.}
Invoking a same Web service several times may help gather additional data for the purpose of
cardinality mediation. However, some Web services cannot be indefinitely invoked when participating in a composition, for several reasons such as cost of the invocation, real world modifications, change of the Web service state, etc. For example, ``add-to-cart'' or ``pay'' operations of a shopping Web service must be invoked exactly once, as their executions make changes in the real world like updating a customer's banking account. On the contrary, some Web services can be indefinitely invoked without any changes in the environment, such as random number generator Web services.

According to the characteristics detailed above, we provide a definition of Web services that includes all the relevant aspects to cardinality mediation:
\begin{definition}[Web service]
A Web service $WS$ along with respective cardinality constraints is
defined as a tuple $\langle ws, CS_{In}, CS_{Out}, Inv_{max}
\rangle$ where,
\end{definition}

\begin{itemize}
\item $ws$ is the Web service's identifier.

\item $CS_{In}$ and $CS_{Out}$ are the constrained schemas that
define the schema and cardinality constraints on data input and data
output of~$ws$.

\item $Inv_{max}$ is the maximum number of allowed invocations in
the composition.
\end{itemize}

For the purpose of cardinality mediation, we noticed a specific category of Web services that are indefinitely invocable and provide new data on each invocation. These Web services present interesting characteristics for
cardinality mediation as they allow gathering new data on each
invocation and thus they can to a certain extent fulfill the
cardinality constraints of other Web services in case of lack of
data. We qualify these Web services as \textit{data providers}. Data provider Web
services include: (i)~biological Web service that sends pieces of
DNA~information, (ii)~mathematical Web services that generate random
numbers, (iii)~Web services that give up-to-date information on a
patient (heartbeat, blood pressure,~etc.), (iv)~geographical
localization Web services, (v)~weather Web services,~etc.

\subsection{Constrained composition representation}
A constrained composition is represented as a combination of Web services and
data flows with shared constraints. Indeed, two data flows connecting to the same Web service share the constraint on its maximum number of invocations.


\begin{definition}[Constrained Web services composition]
A constrained Web services composition $C$ is defined as a set of Web
services $WS$ and data flows $df$ where a data flow is defined as a
tuple $\langle WS_s, WS_r, dup, sel, ord \rangle$, $WS_s$ is the
sender Web service, $WS_r$ is the receiver Web service, $dup$
expresses the tolerance to duplicates, $sel$ expresses the
possibility to select data in the data flow, and $ord$ denotes
whether the order of data elements must be conserved.
\end{definition}

\begin{figure}[ht]
 \centering
 \includegraphics[width=.9\textwidth]{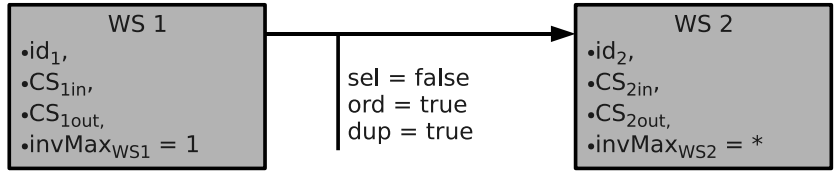}
 \caption{Editor $WS_1$ sending out documents to printing $WS_2$}
 \label{fig:composition_graph}
\end{figure}

To graphically model a constrained Web services composition and facilitate cardinality mediation, a labeled directed graph representation with property sets is adopted~(Fig.~\ref{fig:composition_graph}). In this graph, Web services and data flows correspond to nodes and edges, respectively. A Web service is represented by its constrained input and output schemas and a property that describes the maximum number of times it can be invoked in the composition. An edge has three constraints respectively related to data selection, duplicates and ordering. In Fig.~\ref{fig:composition_graph} a simple data flow between $WS_1$ and $WS_2$ is represented. The properties in this data flow are: data selection is not allowed, data ordering must be conserved, and data duplicates are tolerated. Properties associated with Web services are as follows: Editor $WS_1$ can be invoked at most once, and printing $WS_2$ can be invoked as many times as required.

\section{Classification of cardinality heterogeneities}
\label{sec:problem}

Let us consider the composition of Fig.~\ref{fig:composition_graph}.
$WS_1$ generates a list of elements that comply with a constrained schema $CS_{1out} = (ET, R, s, t, PS_{1out})$ and $WS_2$ requires the reception of elements that comply with a constrained schema $CS_{2in} = (ET, R, s, t, PS_{2in})$. In such a composition, cardinality constraints compatibility consists in checking out if the constraints $PS_{1out}$ and $PS_{2in}$ are compatible\footnote{We remind the reader our work is limited to one couple of constraints per schema for simplicity purpose.}.

Given two cardinality constraints $k_{r1} = [i, j] \in PS_{1out}$ and
$k_{r2} = [m, n] \in PS_{2in}$, respectively associated with $CS_{1out}$ and $CS_{2in}$, different correspondence cases can be identified as shown below:

\begin{figure}[ht]
     \centering
    \includegraphics[width=.6\textwidth]{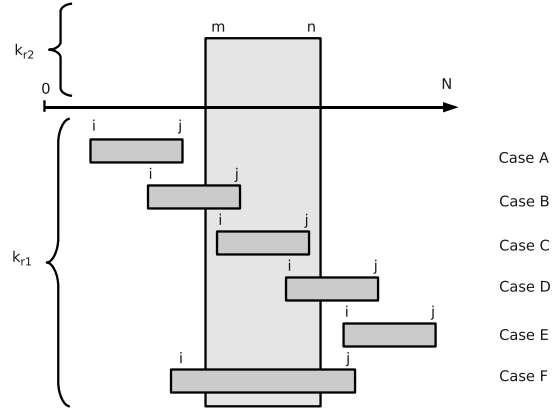}
     \caption{Cardinality constraints compatibility cases\label{fig:cases}}
 \end{figure}

\begin{description}
\item[Case a.] $j < m$: Guaranteed lack of elements. To be
executed, $WS_2$ needs at least $m$ elements from $WS_1$. However,
$WS_1$ can return at most $j$ elements, which is less than needed.
As a result, there is a lack of required elements to make a
$WS_2$ invocation possible.

\item[Case b.]  $i < m \wedge m \leq j \leq n$: Potential lack of
elements. A lack of elements will only occur if, at runtime, the number
of data instances in the $WS_1$ result is smaller than the minimum number
of elements expected by $WS_2$ as an input. Otherwise, cardinality
constraints $PS_{1out}$ and $PS_{2in}$ are compatible.

\item[Case c.] $i \geq m \wedge j \leq n$: No cardinality
constraints compatibility.

\item[Case d.] $m \leq i \leq n \wedge j > n$: Potential overabundance.
When invoked, the Web service $WS_2$ might receive more elements
than needed. There is therefore a potential overabundance of
elements when $WS_2$ is invoked.

\item[Case e.] $i > n$: Guaranteed overabundance. When invoked,
the Web service $WS_2$ will receive more elements than needed. There
is therefore a guaranteed overabundance of elements when $WS_2$ is
invoked.

\item[Case f.] $i < m \wedge j > n$: Potential lack and
overabundance. Depending on the real number of elements of the
$WS_1$ result, a lack or overabundance of elements might occur as explained
previously.
\end{description}

Total satisfaction of constrained schema compatibility only happens in case c. However, in the other cases, it is still possible to reconcile cardinality constraints of Web services by applying appropriate mediation strategies. We remind that cardinality mediation occurs at the instance level. Then, we can group different schema-level heterogeneities into common mediation cases.

\begin{itemize}
\item Lack of elements: cases a and b, possibly case f.
\item Overabundance of elements: cases d and e, possibly case f.
\end{itemize}

We note that case f, depending on the actual number of instances sent, may belong to the ``lack of elements'' situation, to the ``overabundance of elements'' situation.

\section{Cardinality mediation for Web services}
\label{sec:card_mediation}

In this section, we rely on the theoretical framework developed previously and propose a solution based on constraint logic to handle data cardinality in Web services composition. First, we describe the requirements of cardinality mediation for a data flow, and show how these requirements apply by extension to a composition~; second, we quickly introduce constraint logic and show its relevance for the purpose of describing the requirements of cardinality mediation~; and third we present the sets of constraints we developed and show their applicability with a graphical data flow simulation implementation.

\subsection{Requirements of cardinality mediation}

Cardinality mediation requires complex computations in order to adapt data flows between Web services. In the following, we formally describe the requirements for obtaining successful cardinality mediation in a data flow. We identify different situations depending on the \textit{duplicate tolerant} and \textit{data selection tolerant} constraints on the data flow, and explain how the ordering constraint is dealt with.

In order to demonstrate the requirements of cardinality mediation, we consider a data flow between two Web services $WS_1$ and $WS_2$, with structurally matching constrained schemas $CS_{1out}$ and $CS_{2in}$. $CS_{1out}$ holds a constraint $k_{r1} = [a, b]$ and $CS_{2in}$ holds a constraint $k_{r2} = [x, y]$.
We also define the number of invocations $m$ and $n$ of $WS_1$ and $WS_2$ respectively. $[a, b]$ and $[x, y]$ are intervals that represent the possible number of instances that can be obtained on one invocation of $WS_1$ and $WS_2$ respectively, and the operator $*$ applies to these intervals as follows: $m * [a, b]$ is equivalent to $[m*a, m*b]$ to represent the number of instances that can be obtained after $m$ invocations of a Web service.

\subsubsection{Duplicate tolerant/data selection untolerant data flows.}

Let us consider that the aforementioned data flow is duplicate tolerant and data selection untolerant, with $k_{r1} = [9, 11]$ and $k_{r2} = [6, 8]$\footnote{Such cardinality constraints are improbable but they illustrate the complexity of cardinality mediation and show that our solution is applicable to any couples of constraints.}. At first sight, $k_{r1}$ cannot meet the cardinality constraints of $k_{r2}$. Indeed, a first call to $WS_1$ binds $k_{r1}$ to $[9, 11]$ and a second call to $WS_1$ binds $k_{r1}$ to $[18, 22]$, and both do not match $k_{r2}$.

However, we notice that three invocations to $WS_2$ bind $k_{r2}$ to $[18, 24]$. Then, a reconciliation between $WS_1$ and $WS_2$ is possible if $WS_1$ is invoked twice and $WS_2$ is invoked three times, as in this case $k_{r1} \subset  k_{r2}$. Hence, the number of elements required by $WS_2$ is provided by $WS_1$. By extrapolation, we devise when a cardinality mediation for this type of data flow is probably successful when:
\begin{center}
$\exists m, n \in (\mathbb{N}^+)^2$ such that $(m \ast [a, b]) \cap (n \ast [x,y]) \neq \emptyset$.
\end{center}
This situation becomes more and more unlikely to happen as $n$ grows and as $(m \ast [a, b]) \cap n \ast [x,y]$ becomes small.
Accordingly, cardinality mediation for this type of data flow is certain to be successful when:
\begin{center}
$\exists m, n \in (\mathbb{N}^+)^2$ such that $(m \ast [a, b]) \subseteq (n \ast [x,y])$,
\end{center}
which means that $m$ and $n$ verify the following condition: $\dfrac{x}{a} \leqslant \dfrac{m}{n} \leqslant \dfrac{y}{b}$. We remind that such condition applies to duplicate tolerant and data selection untolerant data flows only.

\subsubsection{Duplicate tolerant/data selection tolerant data flows.}

Duplicate and data selection tolerant data flows need calling $WS_1$ as many times as required to exceeding the minimal cardinality required by $WS_2$, and then select elements depending on the users' selection policy (the \textit{``select''} operation is detailed below). The formal representation is trivial and is:

\begin{center}
$\exists m \in \mathbb{N}^+$ such that $(m \ast a) \geqslant x$,
\end{center}

\subsubsection{Duplicate untolerant/data selection untolerant data flows.}

In duplicate untolerant data flows the number of duplicates in a message part is undetermined. Hence, the number of unique elements contained in a message part may vary between $0$ and the maximum number of elements.
On a single run of $WS_1$ , the number of unique elements contained in $CS_{1out}$ is bound between $0$ and $n \ast b$. As duplicate detection and removal applies to the instance level, it is not possible to determine a priori whether or not cardinality mediation will be successful. However, it is possible to describe it at runtime.

Being given $i$ the number of data instances returned by $WS_1$ and a function $f$ that remove duplicates, cardinality mediation for a data selection untolerant, duplicate untolerant data flow successful when:

\begin{center}
$\exists n \in \mathbb{N}^+$ such that $f(i) \in (n \ast [x,y])$.
\end{center}

\subsubsection{Duplicate untolerant/data selection tolerant data flows.}

Accordingly, if data selection is allowed, then cardinality mediation is successful when:

\begin{center}
$\exists n \in \mathbb{N}^+$ such that $f(i) \geqslant (n \ast x)$.
\end{center}

\subsubsection{Ordering on data flows.}

It is not possible to determine order-change tolerance without the intervention of the composition designer. Then, the data ordering is left to the user via an user interface that interacts with the user if necessary. If the data flow supports unordered lists, the cardinality mediator simply concatenates data elements. If ordering is important, alternative strategies are proposed to the user (concatenation of results, mixing of results depending on an algorithm, or manual ordering).

\subsubsection{Application to a composition.}

In this section, we presented the requirements of a composition for one data flow. Indeed, these requirements can be scaled up to a composition. In such case, numbers of invocations $m$, $n$, cardinality constraints $[a, b]$, $[x, y]$ are shared between several data flows. Such situations reduces the number of possibilities of the composition to succeed, however it simplifies the resolution of cardinality requirements as it provides a unified view of the whole business process with all the cardinality constraints.

\subsection{Constraint logic for cardinality mediation}

It turns out that constraint logic programming is well-suited for modelling cardinality mediation.
%
%
More precisely, constraint logic programming over finite domain variables
allows to specify constraints over these variables and to use these
constraints in an a priori way to reduce the search space. The
resulting framework is quite elegant since, on the one hand, it
conserves the declarative expression of logic programming, including
the multi-directionalities of its queries, and, on the other hand, it
integrates an efficient way of solving constraints.

This is very appealing in our cardinality mediation context. For
instance, the following predicate
$basic\_{}mediation(A,B,X,Y,M,N,D,S)$
aims at determining whether
there are $M$, $N$ such that M * [A,B] is a subset of N * [N,M] for a duplicate tolerant and data selection untolerant data flow. Such predicate describes the constraints we devised:
\begin{verbatim}
basic_mediation(A,B,X,Y,M,N,Mmax,Nmax,D,S) :-

   fd_domain([M],1,Mmax),
   fd_domain([N],1,Nmax),
   N * X #=<# M * A,
   M * B #=<# N * Y,
   fd_labeling([M,N]).
\end{verbatim} 

The couples $A, B$ and $X, Y$ represent the cardinality constraints of respectively $WS_1$ and $WS_2$, $M, N$ are the number of invocations of $WS_1$ and $WS_2$ to be found, $Mmax$ and $Nmax$ their maximum number of possible invocations, $D$ and $S$ are booleans that describe duplicate tolerance and data selection respectively.

 The code first defines the intervals [1..Mmax] and [1..Nmax] as finite domains for the
 variables $M$ and $N$ and then specifies the constraints as given
 before, with the symbols \verb|#=<#| indicating the less than equal
 relation. Finally, the $fd\_{}labeling$ predicate is used to start an
 exhaustive search but by first fixing a value for M, then by propagating
 this value in the constraints to reduce the domain of $N$ and finally
by searching in the reduced domain $N$ for a value.

An example is worth to capture what this means.
Let us consider two Web services that hold [9, 11] and [6, 8] as cardinality constraints.
The corresponding query is $basic\_{}mediation(9, 11, 6, 8, M, N, 10, 10, true, false)$.
Thus the domains of $M$ and $N$ are both limited to [1..10]. Then $M$ is fixed to $1$ and the constraints become $N * 6 \leqslant 9$ and $11 \geqslant N * 8$. All the possible values of $N$ are then searched but there is no integer value of $N$ that is comprised between $11/8$ and $9/6$. Then $M$ is fixed to $2$ and the constraints become $N * 6 \leqslant 18$ and $22 \geqslant N * 8$. In such case, $N = 3$ is comprised between $22/8$ and $18/6$. Hence, the first solution returned by the Prolog engine is the couple of values $(M = 2, N = 3)$. Other solutions are possible but they require additional invocations of $WS_1$ or $WS_2$, thus the first solution is the most interesting one.

\subsection{Implementation}

\begin{figure}[ht]
 \centering
 \includegraphics[width=.4\textwidth]{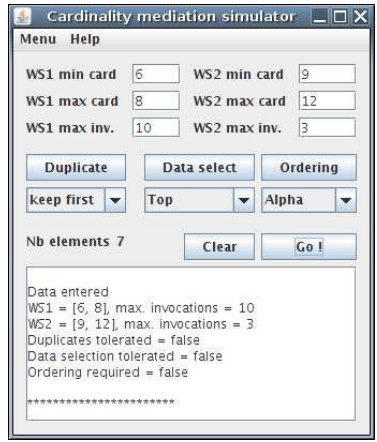}
 \caption{Screenshot of the data cardinality mediation simulator}
 \label{fig:screenshot}
\end{figure}

We developed a proof-of-concept prototype tool that simulates a data flow between Web services and shows the possibilities offered by our cardinality mediation approach (Figure~\ref{fig:screenshot}). This prototype tool relies on Java platform, and connects to a Prolog reasoner that contains cardinality mediation functions encoded as shown in the above. Our Java program is a client application and the Prolog engine is currently deployed in a Web server as a CGI program that relies on GNU Prolog~\cite{DBLP:journals/jflp/DiazC01}. The Java program is actually a CGI client and user frontend, its performs the CGI calls and interacts with the end-user. The mediation process is performed by the Prolog engine, it depends on the constraints on Web services and data flow entered via the interface.

Our solution operates as follows: for each data flow, the end-user feeds a Prolog engine with the different constraints on Web services and data flow via our graphical interface. Then, the Prolog engine calculates the possibilities to obtain a successful result and selects the best result if found.

The constraints are \textit{a priori} required and the user needs to provide them before runtime. These are the constraints on the maximum number of executions for $WS_1$ and $WS_2$, together with the constraints on data flow (Data selection, duplicates and ordering), which are described neither in typical (WS-BPEL) business processes nor in typical Web service descriptions (WSDL). We are currently working on a Web service composition platform that connects to our cardinality mediation simulator and feeds it with the composition parameters, in order to apply our constraint-based reasoning to the whole composition.


Cardinality mediation involves different operations on data flows. We defined the following operations:
\begin{itemize}
\item $select(O, strategy)$ selects particular elements of the result set $O$ depending on a user selection strategy (first elements, each two elements, last elements, manual selection\dots)

\item $merge(O_{1}, O_{2}, strategy)$ merges lists of results $O_{1}$ and $O_{2}$ depending on a user merging strategy (concatenate elements of the second call to the first, mix each two elements, concatenate elements of the first call to the second, manual merging\dots)

\item $rm\_dup(O, strategy)$ removes all the duplicates in $O$ according to a user selection strategy (remove first duplicate first, remove last duplicate first).
\end{itemize}
These operations are performed with the help of the user via the graphical interface.

\section{Conclusion}
\label{sec:conclusion}

In this paper, we shed the light on an important issue that could refrain
the smooth progress of service composition scenarios if not addressed
properly. This issue, which we referred to as cardinality
heterogeneity, stresses out the importance of quantifying the data
that Web services engaged in these scenarios exchange.
Little research work has been done to
address this issue. In this paper, we proposed a classification of cardinality
heterogeneities and highlighted
for example how a Web service could be overloaded with data that it
might not need, and how lack of expected data could degrade
the Web service compostition. A  constraint logic-based  cardinality mediation approach
is proposed.  It aims to adapt data flows between Web services by considering
the requirements of both data flows in terms of tolerance to duplicate, selection and ordering, and Web services in terms of number of allowed invocations and constrained schemas.

As a future work, we intend to study how different heterogeneous web services providing a same
functionality could be combined to cover the cardinality constraint of a service.
A Web services community-based approach could be adopted.
We are also interested in handling data cardinality of web services in the context
of data privacy where some data might be hidden or shown depending on the
existing privacy policies.

\bibliographystyle{abbrv}
\bibliography{ref}

\end{document}